# A Multilayer Model of Computer Networks

Andrey A. Shchurov

*Department of Telecommunications Engineering, Faculty of Electrical Engineering*
*Czech Technical University in Prague, Czech Republic*

***Abstract***— *The fundamental concept of applying the system methodology to network analysis declares that network architecture should take into account services and applications which this network provides and supports. This work introduces a formal model of computer networks on the basis of the hierarchical multilayer networks. In turn, individual layers are represented as multiplex networks. The concept of layered networks provides conditions of top-down consistency of the model. Next, we determined the necessary set of layers for network architecture representation with regard to applying the system methodology to network analysis.*

## I. INTRODUCTION

Applying a system methodology to network analysis [1] is a relatively new approach, particularly in the Internet Protocol (IP) world. The fundamental concept is that network architecture should take into account services and applications which this network provides and supports. It is important to note that this concept is completely supported by the most recent practical approaches such as Business-Driven Design [2] and Application Centric Design [3].

On the other hand, one of the major goals of modern physics is providing proper and suitable representations of systems with many interdependent components, which, in turn, might interact through many different channels. As a result, interdisciplinary efforts of the last fifteen years have leaded to the birth of complex networks theory [4] which uses graphs as powerful mathematical tools for modelling pairwise relationships among sets of objects/entities. Traditionally graphs capture only a single form of relationships between objects [5]. However, complex networks rely on different forms of such relationships, which can be naturally represented by multilayer networks [6]. Assuming that all layers are informative, they can provide complementary information. Thus, we can expect that a proper combination of the information contained in the different graph layers leads to a formal network representation (a formal model) which will be appropriate for applying the system methodology to network analysis.

The rest of this paper is structured as follows. Section 2 introduces the related work. Section 3 presents the formal model of computer networks based on the concept of multilayer networks. In turn, Section 4 focuses on the set of layers for network architecture representation. Finally, conclusion remarks are given in Section 5.

## II. RELATED WORK

The multilayer approach for the modelling of computer networks can be roughly classified into two categories:
- multilayer networks (network architecture representation);
- reference models (layers definition).

### A. Multilayer networks

Recent surveys in the domain of multilayer networks provided by Kivela et al. [7] and Boccaletti et al. [8] give a comprehensive overview of the existing technical literature and summarize the properties of various multilayer structures. However, it is important to note that the terminology referring to systems with multiple different relations has not yet reached a consensus – different papers from various areas represent similar terminologies to refer to different models, or distinct names for the same model.

### B. Reference models

The ISO/OSI Reference Model (OSI RM) [9] was developed years ago for application developers, equipment manufacturers and network protocol vendors as an open standard for constructing network devices and applications/services that can work together. The model partitions computing systems into seven abstraction layers:

1. Physical Layer;
2. Data Link Layer;
3. Network Layer;
4. Transport Layer;
5. Session Layer;
6. Presentation Layer;
7. Application Layer.

However, this conceptual model has never been implemented in practice. Instead of it, increasing popularity of TCP/IP based networking has led developers to use the TCP/IP Protocol Suit [10] [11] [12] which five layers are based on OSI RM – layers 5 through 7 are collapsed into the Application Layer.

In turn, IETF standards RFC 1122 [13] and RFC 1123 [14] define the TCP/IP Reference Model. This model is compatible with OSI RM and TCP/IP Protocol Suit but it partitions computing systems into four abstraction layers – layer 1 (Physical Layer) is removed from the model.

### III. MULTILAYER MODEL

A type of multilayer network of particular relevance for computer networks is a *hierarchical multilayer network* [7], in which the bottom layer constitutes a *physical* network and the remaining layers are *virtual layers* that operate on top of the physical layer. On the other hand, the concept of layered networks [15] provides the condition of top-down consistency based on the facts:
- for each node on a given layer there is a corresponding node (or nodes) on the layer below;
- for each logical path between two nodes on a given layer there is a path (or paths) between the corresponding nodes on the layer below.

Thus, the formal basic definitions of multilayer networks [8] (adapted to the hierarchical top-down approach) can be used as a starting point. In this case:

**Definition 1:** *Let the graph $M = (V, E)$ denote the computer network as a hierarchical multilayer network:*

$$M = (V, E)$$

*where $M$ is a multi-layered 3D graph (see Fig. 1); $V(M)$ is a finite, non-empty set of components (hardware or software); and $E(M)$ is a finite, non-empty set of component-to-component interconnections (information links). In turn:*

$$V = \bigcup_{\alpha=1}^{N} V_\alpha$$

*and*

$$E = \left(\bigcup_{\alpha=1}^{N} E_\alpha\right) \bigcup \left(\bigcup_{\alpha=2}^{N} E_{\alpha,(\alpha-1)}\right)$$

*where $V_\alpha$ is a finite, non-empty set of components on layer $\alpha$; $E_\alpha$ is a finite, non-empty set of intralayer component-to-component interconnection on layer $\alpha$; $E_{\alpha,(\alpha-1)}$ is a finite, non-empty set of interlayer relations (projections) between components of a given layer $\alpha$ and the layer below $(\alpha - 1)$; and $N$ is the number of layers.*

On the other hand, two main elements of multilayer networks are [7]:
- intra-layer graphs;
- inter-layer graphs.

In this case:

**Definition 2:** *Let the subgraph $G_\alpha$ denote a layer of the computer network:*

$$G_\alpha = (V_\alpha, E_\alpha)$$

*where $G_\alpha$ is an intralayer subgraph of $M$; $V_\alpha$ is a finite, non-empty set of components on layer $\alpha$; and $E_\alpha \subseteq V_\alpha \times V_\alpha$ is a finite, non-empty set of intralayer component-to-component interconnections on layer $\alpha$.*

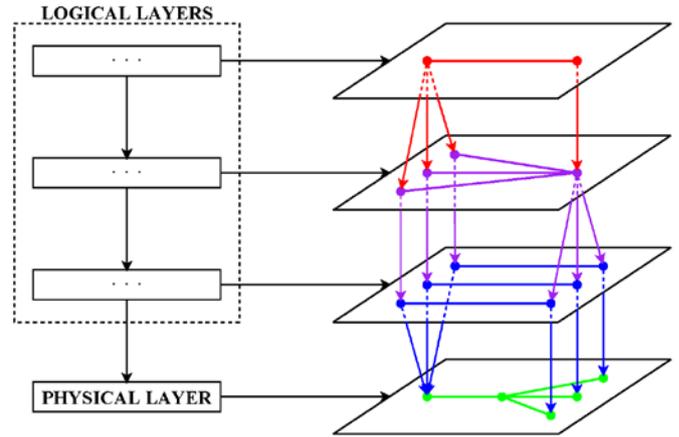

Fig. 1 Multilayer model of computer networks.

**Definition 3:** *Let the subgraph $G_{\alpha(\alpha-1)}$ denote a cross-layer of the computer network:*

$$G_{\alpha(\alpha-1)} = (V_\alpha, V_{(\alpha-1)}, E_{\alpha,(\alpha-1)})$$

*where $G_{\alpha,(\alpha-1)}$ is an interlayer bipartite subgraph of $M$; $V_\alpha$ is a finite, non-empty set of components on layer $\alpha$, $V_{(\alpha-1)}$ is a finite, non-empty set of components on layer $(\alpha - 1)$; $E_{\alpha,(\alpha-1)} \subseteq V_\alpha \times V_{(\alpha-1)}$ is a finite, non-empty set of interlayer relations (projections) between components of a given layer $\alpha$ and the layer below $(\alpha - 1)$.*

As mentioned above, the concept of layered networks [15] strictly relies on the fact that a node in a given layer depends on a corresponding node in the layer below, i.e. each individual component $v_i^\alpha \in V_\alpha$ has at least one corresponding neighbour $v_j^{(\alpha-1)} \in V_{(\alpha-1)}$ where $v_j^{(\alpha-1)}$ must satisfy the condition $(v_i^\alpha, v_j^{(\alpha-1)}) \in E_{\alpha,(\alpha-1)}$, i.e. $|V_\alpha| \leq |E_{\alpha,(\alpha-1)}|$.

It is important to note that the degree (or valency) of vertices of $G_{\alpha,(\alpha-1)}$ represents the technological solutions which were used to build the system [16]:
- $d(v_i^\alpha) > 1$; $v_i^\alpha \in V_\alpha$ – clustering technology representation;
- $d(v_j^{(\alpha-1)}) > 1$; $v_j^{(\alpha-1)} \in V_{(\alpha-1)}$ – virtualization and replication technologies representation;
- $d(v_i^\alpha) = d(v_j^{(\alpha-1)}) = 1$ – a special case of dedicated components.

Based on Definitions 2 and 3, computer networks can be represented as:

$$M = \left(\bigcup_{\alpha=1}^{N} G_\alpha\right) \bigcup \left(\bigcup_{\alpha=2}^{N} G_{\alpha,(\alpha-1)}\right)$$

From the perspective of system methodology, intralayer subgraphs $G_\alpha$ are the main source of initial data for the network analysis processes; and interlayer subgraphs $G_{\alpha,(\alpha-1)}$ make these processes consistent on all layers of the formal model.

In practice, intralayer subgraphs $G_\alpha$ are not monolithic structures: a set of protocols is predefined for each (physical or virtual) layer. For example, a wireless access point (AP) must support at least two different protocols: (1) one for wired and (2) one for wireless communications. Moreover, these protocols can support different topologies. As a consequence, each intralayer subgraph $G_\alpha$ consists of a fixed set of components connected by different types of information links.

A type of multilayer network of particular relevance for this case are *multiplex* [6] [17] or *multidimensional* [18] networks, in which different layers represent different types of component-to-component interconnections. Hence, we should rewrite Definition 2 as follow:

**Definition 2\*:** *Let the subgraph $G_\alpha = (V_\alpha, E_\alpha, S_\alpha, P_\alpha)$ denote a layer of the computer network, where $G_\alpha$ is a labeled intralayer subgraph of $M$; $V_\alpha$ is a finite, non-empty set of components on layer $\alpha$; $E_\alpha$ is a finite, non-empty set of intralayer component-to-component interconnections on layer $\alpha$; $S_\alpha$ is a vertices label set; and $P_\alpha$ is a set of predefined protocols for layer α. In this case:*

$$S_\alpha = \bigcup_{v_i^\alpha \in V_\alpha} S_i^\alpha$$

where $S_i^\alpha$ is a finite non-empty totally ordered set of component specifications (at least the set of supported communication protocols) or the label of the vertex $v_i^\alpha$ of $G_\alpha$. In turn, $G_\alpha$ is represented as a multiplex network (see Fig. 2):

$$G_\alpha = \bigcup_{\beta=1}^{|P_\alpha|} G_\beta^\alpha$$

where $G_\beta^\alpha = (V_\alpha, E_\beta^\alpha)$ is a sub-subgraph which is defined by the communication protocol $p_\beta^\alpha \in P_\alpha$; and $E_\beta^\alpha \subseteq E_\alpha$ is a finite, non-empty set of intralayer component-to-component interconnections on sub-layer β of layer α. An edge $(v_i^\alpha, v_j^\alpha) \in E_\alpha$ belongs to $G_\beta^\alpha$ iff both components $v_i^\alpha$ and $v_j^\alpha$ support this protocol, i.e. $p_\beta^\alpha \in S_i^\alpha$ and $p_\beta^\alpha \in S_j^\alpha$ (each pair of components $v_i^\alpha$ and $v_j^\alpha$ can be connected by at most $|P_\alpha|$ possible edges).

Based on Definitions 2\* and 3, computer networks can be finally represented as:

$$M = \left(\bigcup_{\alpha=1}^{N}\left(\bigcup_{\beta=1}^{|P_\alpha|} G_\beta^\alpha\right)\right) \bigcup \left(\bigcup_{\alpha=2}^{N} G_{\alpha,(\alpha-1)}\right)$$

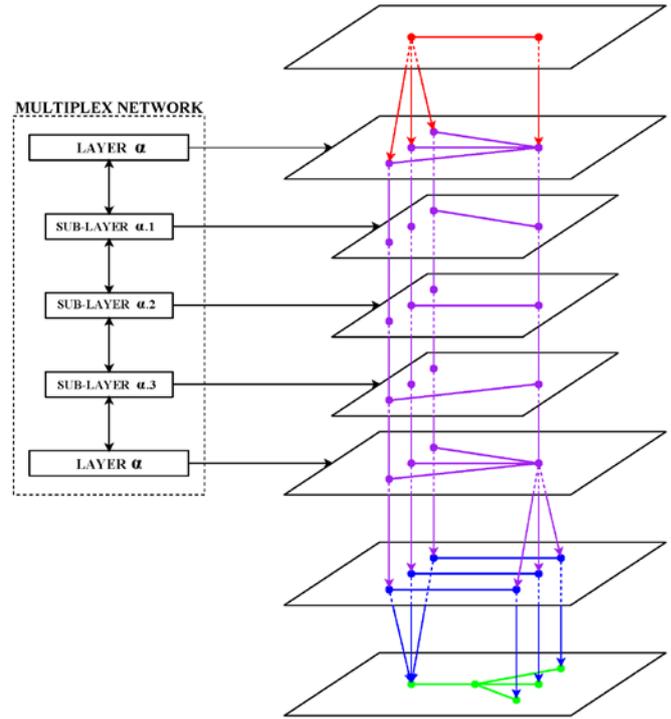

Fig. 2 Intralayer subgraph representation as a multiplex network.

### IV. LAYERS DEFINITION

As mentioned above, the ISO/OSI Reference Model has never been implemented in practice. Thus, the TCP/IP Protocol Suit can be used as a starting point. On the other hand, network architecture representation should be as clean and simple to understand as it can be. As a consequence (in contrast to the developer community) the business community (end-users) faces the following challenges [16]:

- Physical Layer and Data Link Layer cannot be divided in the case of commercial off-the-shelf (COTS) network equipment;
- Transport Layer and Application Layer cannot be divided in the case of COTS software.

Moreover, end-users do not need services and applications themselves – they need tools to solve their business problems. However, neither OSI RM not TCP/IP Protocol Suit provides a layer to represent the enlarged viewpoint of end-users (business goals). As a consequence, a common joke is that OSI RM should have three additional layers [19]:

8. User Layer;
9. Financial Layer;
10. Political Layer.

In practice computer network focus on solving problems at layer 10 (but they are usually limited by layer 9).

Hence, we should define the additional layer which can represent the enlarged system functionality or business goals.

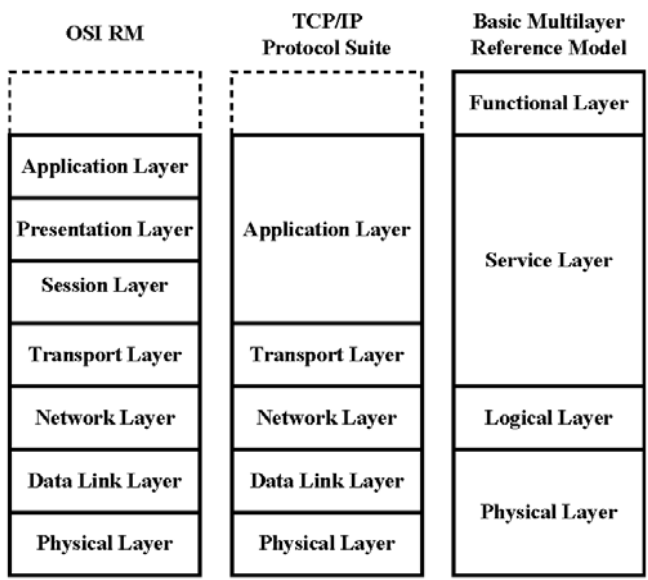

Fig. 3 Basic multilayer reference model.

As a consequence, this layer is based on functional models [1] (end-user requirements representation).

The basic multilayer reference model is shown in Fig. 3. From the viewpoint of the hierarchical multilayer network, the physical layer constitutes a physical network and the logical, service and functional layers are virtual layers that operate on top of the physical layer.

Unfortunately, the basic model does not take into account the environment impact. The problem can be solved by two additional layers [20] (see Fig. 4):

- *The engineering environment layer*. This layer defines external engineering systems (power supply systems, climate control systems, physical security systems, etc.) that are vital for normal operation of computer networks. It is based on topological models [1], where engineering systems and computer networks are represented as individual components.
- *The social environment layer* (or layer 8 of OSI RM [19]). This layer defines organization infrastructures or human networks. It represents persons or groups of persons and their working relationships based on electronic communications.

It is important to note that all these additional layers lie beyond the ISO/OSI RM and the TCP/IP Protocol Suit but they provide a necessary complement to it with regard to applying the system methodology to network analysis.

## V. CONCLUSIONS

The fundamental concept of applying the system methodology to network analysis is that network architecture should take into account services and applications which this network provides and supports.

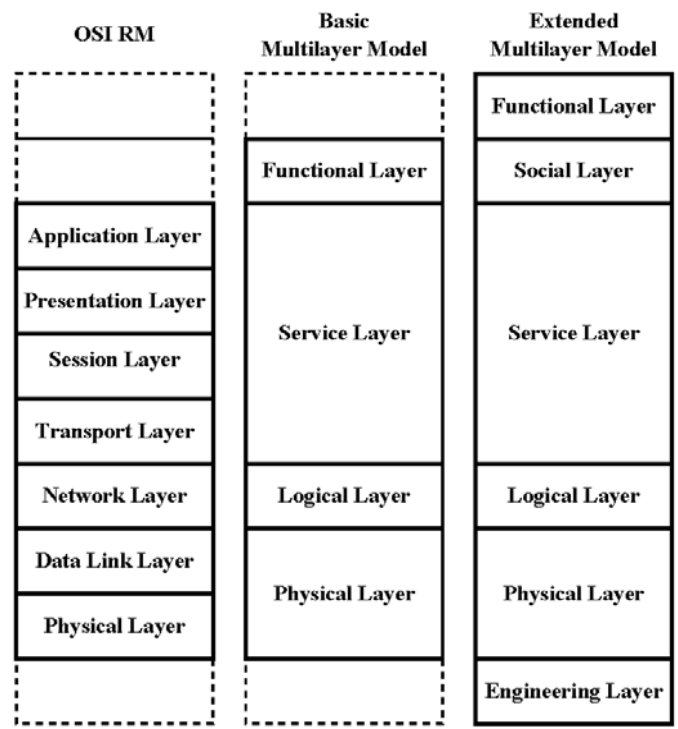

Fig. 4 Extended multilayer reference model.

In this work we determined a formal model of computer networks on the basis of the hierarchical multilayer networks in which the bottom layer constitutes a physical network and the remaining layers are virtual layers that operate on top of the physical layer. In turn, the representation of individual layers is based on the concept of multiplex networks in which different sub-layers represent different types of component-to-component interconnections for a fixed set of components (hardware or software). On the other hand, the concept of layered networks provides the condition of top-down consistency of the model.

Next, we determined the necessary set of layers for network architecture representation with regard to applying the system methodology to network analysis. This set covers:

- hardware-based aspects (components and their interconnections) of computer networks;
- software-based aspects;
- network environments (external engineering systems and organization infrastructures);
- network business goals based on end-user requirements

Using this model and the graph theoretical metrics, both static and dynamic system analysis can be performed:

- The static analysis determined the characteristics of each layer based on the layer structure (or topology) [21] [22].
- The dynamic analysis (or fault injection simulation) provides a means for understanding how distributed systems behave in the presence of faults [23] [24].

In turn, the approaches of automated transformation of network specifications and end-user requirements into abstract formal models are beyond the scope of this paper. This problem requires a separate analysis – in the case of complex or non-standard computer networks, it may not be a routine exercise in practice.

ACKNOWLEDGMENT

This article has originated within the framework of research and development activities at the Department of Telecommunication Engineering (Czech Technical University in Prague, Faculty of Electrical Engineering).